\documentclass[a4paper,twocolumn,prl,showpacs]{revtex4}
\renewcommand{\[}{\begin{equation}}
\renewcommand{\]}{\end{equation}}
\newcommand{\equ}[1]{Eq.~(\ref{#1})}
\newcommand{\eqs}[2]{Eqs.~(\ref{#1}) and (\ref{#2})}
\def\rmp#1#2#3{Rev. Mod. Phys. {\bf #1}, #2 (19#3)}

\newcommand{\ei}[1]{{\rm e}^{i #1}}

\newcommand{\intr}{\int \! d{\bf r} \;}

\def\runtime{(\the\time)\qquad\the\month/\the\day/\the\year}% get current time
\def\today
 {\count10=\year\advance\count10 by -1900 \number\day--\ifcase
  \month \or Jan\or Feb\or Mar\or Apr\or May\or Jun\or
             Jul\or Aug\or Sep\or Oct\or Nov\or Dec\fi--\number\count10}
\def\hour{\count10=\time\count11=\count10
\divide\count10 by 60 \count12=\count10
\multiply\count12 by 60 \advance\count11 by -\count12\count12=0
\number\count10 :\ifnum\count11 < 10 \number\count12\fi\number\count11}
\def\bea{\begin{eqnarray}}
\def\eea{\end{eqnarray}}
\def\nn{\nonumber\\}
\begin{document}

\title{Polarization fluctuations in insulators and metals: New and old theories merge}

\author{Raffaele Resta}

\affiliation{INFM DEMOCRITOS National Simulation Center, via Beirut 2,
I--34014 Trieste, Italy \\ and Dipartimento di Fisica Teorica, Universit\`a di
Trieste, Strada Costiera 11, I--34014 Trieste, Italy}

\date{}

\begin{abstract} The ground-state fluctuation of polarization {\bf P} is
finite in insulators and divergent in metals, owing to the SWM sum rule [I.
Souza, T.  Wilkens, and R. M. Martin, Phys. Rev. B {\bf 62}, 1666 (2000)].
This is a virtue of periodic (i.e. transverse) boundary conditions. I show
that within any other boundary conditions the {\bf P} fluctuation is finite
even in metals, and a generalized sum rule applies. The boundary-condition
dependence is a pure correlation effect, not present at the
independent-particle level. In the longitudinal case $\nabla \cdot {\bf P} =
-\rho$, and one equivalently addresses charge fluctuations: the generalized
sum rule reduces then to a well known result of many-body theory.
\end{abstract}


\maketitle \bigskip\bigskip

In a paper appeared in 2000 Souza, Wilkens, and Martin~\cite{Souza00}
(SWM) proved a  fluctuation-dissipation sum rule relating the ground-state
fluctuation of polarization {\bf P} in a quantum system to its macroscopic
conductivity. The sum rule implies that {\bf P} fluctuations are finite in
insulators and divergent in metals, thus providing a clearcut {\it qualitative}
difference between insulating and metallic ground states. In fact SWM complete
the program initiated in 1964 by W. Kohn with his ``Theory of the
insulating state''~\cite{Kohn64}. However, a different and apparently unrelated
fluctuation-dissipation sum rule is well known  since the 1950s in many-body
physics~\cite{Pines,Noz}. The disturbing fact is that metals and insulators {\it
do not} behave in a qualitatively different way as far as the latter sum rule is
concerned. I show here that both sum rules are special cases of a more general
one, the difference owing to the boundary conditions (BCs) adopted when taking
the thermodynamic limit: SWM adopt periodic Born--von--K\`arm\`an 
BCs, i.e. transverse, while within many-body physics it is
customary to adopt longitudinal ones. The precise meaning of ``transverse'' and
``longitudinal'' in the present context is illustrated below. For any BC choice
different from the purely  transverse the ground-state {\bf P} fluctuation is
finite even in metals. I also show that the BC dependence of the
{\bf P} fluctuation is a combined effect of the long range of Coulomb
interaction and of {\it electron correlation}. There is no such dependence for
independent electrons (either Hartree-Fock or Kohn-Sham), where the standard SWM
sum rule applies anyhow. The presentation starts considering a finite system
with ``open'' BCs, and then proceeds to taking the thermodynamic
limit in the appropriate way.
 
Let $\Psi$ be the singlet ground-state wavefunction of an $N$-electron
system, with even $N$,  within ``open'' BCs, i.e., $\Psi$ is
square-integrable. We address the first and second moments of the position
operator \[ \hat{\bf R} = \sum_{i=1}^N {\bf r}_i . \] Since it is expedient
to deal with quantities that become intensive in the large-$N$ limit, I
define the first and second cumulant moments {\it per electron}: \[ \langle
r_\alpha \rangle_{\rm c} = \frac{1}{N} \langle \Psi | \hat{R}_\alpha | \Psi
\rangle = \frac{1}{N} \intr r_\alpha \, n({\bf r}) ; \label{first} \] \[
\langle r_\alpha r_\beta \rangle_{\rm c} = \frac{1}{N} ( \, \langle \Psi |
\hat{R}_\alpha \hat{R}_\beta | \Psi \rangle - \langle \Psi | \hat{R}_\alpha |
\Psi \rangle \langle \Psi | \hat{R}_\beta | \Psi \rangle \, )  \label{second}
\] (Greek subscripts indicate Cartesian components throughout). The first
moment (times the trivial factor $-e N/V$) is the electronic term in the
macroscopic polarization ${\bf P}$, while the second cumulant moment $\langle
r_\alpha r_\beta \rangle_{\rm c}$ is its quadratic quantum fluctuation in the
many-body ground state. We notice that the second moment is a function of the
relative coordinates, while the first moment is a function of the
absolute ones; indeed, the macroscopic polarization ${\bf P}$ is well defined
only when the (classical) nuclear contribution is accounted for. As said
above, SWM address the ${\bf P}$ fluctuations in extended systems within
periodic BCs.  Therein, the position $\hat{\bf R}$ is a ``forbidden''
operator~\cite{rap100} and the definition of $\langle r_\alpha r_\beta
\rangle_{\rm c}$ looks formally quite different from
\equ{second}~\cite{rap107,rap_a23}. 

Indicating with ${\bf x}_i \equiv ({\bf r}_i,\sigma_i)$ the space and spin
coordinates of the $i$-the electron, the one-body and two-body densities are
defined as: \[ n({\bf r}_1) = N \sum_{\sigma_1} \int d {\bf x}_2 \cdots d {\bf
x}_N | \Psi({\bf x}_1,{\bf x}_2,\dots {\bf x}_N) |^2 ; \label{dgen} \] \[
n^{(2)}({\bf r}_1,{\bf r}_2) = N (N-1) \sum_{\sigma_1 \sigma_2} \int d {\bf x}_3
\cdots d {\bf x}_N | \Psi({\bf x}_1,\dots {\bf x}_N) |^2 .  \] Straightforward
manipulations lead to the equivalent form: \bea \langle r_\alpha r_\beta
\rangle_{\rm c} & = & \frac{1}{2N}  \int d {\bf r} d {\bf r}'\;({\bf r} - {\bf
r}')_\alpha ({\bf r} - {\bf r}')_\beta \, \nn &\times& [\,  n({\bf r}) n({\bf
r}') - n^{(2)}({\bf r},{\bf r}') \, ] .\label{bform} \eea We observe that, for
large values of the relative coordinate ${\bf r} - {\bf r}'$, the electron
distribution becomes uncorrelated and one has $n^{(2)}({\bf r},{\bf r}') \simeq
n({\bf r}) n({\bf r}')$: this fact can be regarded as a manifestation of the
``nearsightedness'' principle~\cite{Kohn96}. For any finite $N$ the integration
in \equ{bform} obviously converges, owing to the boundedness of the ground
wavefunction $\Psi$.  One of the main questions is whether  $\langle r_\alpha
r_\beta \rangle_{\rm c}$ goes to a well defined finite limit or instead diverges
in the limit of large $N$.

A well known exact sum rule relates the two-body density to the frequency
integral of the imaginary part of the linear response: such relationship belongs
to the general class of fluctuation-dissipation theorems~\cite{Kubo2,Forster}.
By definition, the linear polarizability tensor $\alpha_{\beta\gamma}(\omega)$
yields the $\beta$ component of the dipole ${\bf d}$ linearly induced by an
electric field ${\bf E}_0$ of unit magnitude in the $\gamma$ direction, at
frequency $\omega$. I address purely electronic response, therefore assuming
{\it clamped nuclei}. Furthermore I stress that ${\bf E}_0$ is the field far
outside the finite sample, different from the screened macroscopic field ${\bf
E}$ inside. Starting e.g from Eq. (2.17) in Ref.~\cite{McLachlan64} and using
\equ{bform}, it is straightforward to prove the sum rule: \[ \frac{1}{N}
\int_0^\infty \!\!\! d \omega \; \mbox{Im } \alpha_{\beta\gamma}(\omega) =
\frac{\pi e^2}{\hbar} \langle r_\beta r_\gamma \rangle_{\rm c} \label{sumrule} .
\] It is worth noticing that the rhs is by construction a ground-state property,
while the lhs is a property of the {\it excitations} of the system. In
\equ{sumrule}, fluctuation and dissipation are perspicuous: from the definition
of \equ{second} the rhs is a quantum fluctuation, while the imaginary part of
$\alpha(\omega)$ measures dissipation in the zero-temperature
limit~\cite{Kubo2}.  Because of the sum rule, the actual value of $\langle
r_\alpha r_\beta \rangle_{\rm c}$ in a given system can in principle be measured
by actually probing the excited states.

I now discuss \equ{sumrule} in the limit a macroscopic solid, where for the sake
of simplicity the bulk is assumed as macroscopically homogeneous and isotropic.
Therefore the macroscopic polarization ${\bf P} = {\bf d}/V$ linearly induced by
a macroscopic field at frequency $\omega$ can be written as: \[ {\bf P} =
\frac{\varepsilon(\omega) - 1}{4 \pi} \, {\bf E}  , \label{epsi} \]   where
$\varepsilon(\omega)$ is the electronic (clamped-nuclei) macroscopic dielectric
function of the bulk material. In terms of ${\bf E}_0$, this polarization is \[
P_\beta = \frac{1}{V} \sum_\gamma \alpha_{\beta\gamma}(\omega) E_{0,\gamma} , \]
where the relationship between ${\bf E}$ and ${\bf E}_0$ depends on the {\it
shape} of the sample. 

Following a well-known practice for the study of dielectric bodies, we consider
a sample of ellipsoidal shape, in which case the ${\bf E}$ field is constant in
the bulk, and the shape effects are embedded in the depolarization coefficients
${n}_\beta$, with $\sum_\beta {n}_\beta = 1$. The main relationship
is~\cite{Landau1}: \[ E_\beta = E_{0,\beta} -4 \pi n_\beta P_\beta .
\label{simple} \] The extremely prolate ellipsoid ($n_x = n_y = 1/2, n_z=0$) is
a cylinder along $z$, while the extremely oblate one ($n_x = n_y = 0, n_z=1$) is
a slab normal to $z$. The slab geometry epitomizes both the longitudinal and the
transverse cases: ${\bf P}$ is purely longitudinal when along $z$, and purely
transverse when along $xy$. In the former case, in fact, we have $P_z = P_z(z)$
(independent of $xy$): hence $\nabla \cdot {\bf P} \neq 0$, $\nabla \times {\bf
P} = 0$. Conversely in the latter case we have $P_x = P_x(z)$ (independent of
$xy$): hence $\nabla \cdot {\bf P} = 0$, $\nabla \times {\bf P} \neq 0$. It is
worth noticing that the charge is uniquely related to ${\bf P}$  via $\nabla
\cdot {\bf P} = -\rho$ in the longitudinal case, whereas the charge {\it does
not enter} a macroscopic description in the transverse one.

In the ellipsoidal geometry \eqs{epsi}{simple} yield~\cite{Landau1}: \[ E_\beta
= \frac{1}{1 + {n}_\beta [ \varepsilon(\omega)  - 1]} E_{0,\beta} . \label{ezer}
\]  \[ P_\beta = \frac{1}{4\pi} \frac{\varepsilon(\omega)  - 1}{1 + {n}_\beta [
\varepsilon(\omega)  - 1]} E_{0,\beta} . \label{ellips} \]  The $\alpha$ tensor
is diagonal over the ellipsoid axes, and the above results transform
\equ{sumrule} into: \[ \delta_{\beta\gamma} \; \frac{V}{4 \pi N} \int_0^\infty
\!\!\! d \omega \; \mbox{Im }  \frac{\varepsilon(\omega)  - 1}{1 + {n}_\beta [
\varepsilon(\omega)  - 1]} = \frac{\pi e^2}{\hbar} \langle r_\beta r_\gamma
\rangle_{\rm c} \label{sumrule2} . \] 

It is expedient to recast this sum rule in terms of the conductivity
$\sigma(\omega)$, which by definition measures the macroscopic current
linearly induced by a field ${\bf E}$ at frequency $\omega$. Since the
current is the time derivative of the electronic polarization, \equ{epsi}
yields $\varepsilon(\omega) - 1 = 4 \pi i \sigma(\omega)/\omega$ and \[
\langle r_\beta r_\gamma \rangle_{\rm c} = \delta_{\beta\gamma} \;
\frac{V}{N} \frac{\hbar}{\pi e^2} \int_0^\infty \frac{d \omega}{\omega} \;
\mbox{Im } \frac{i \sigma(\omega)}{1 + 4 \pi i n_\beta \sigma(\omega)/\omega}
\label{sumrule3} , \] which generalizes the SWM sum rule. In fact the
assumption of periodic BCs corresponds---as I am going to explain below---to
the choice $n_\beta =0$, yielding the original SWM sum rule: \[ \langle
r_\beta r_\gamma \rangle_{\rm c} = \delta_{\beta\gamma} \; \frac{V}{N}
\frac{\hbar}{\pi e^2} \int_0^\infty \frac{d \omega}{\omega} \; \mbox{Re }
\sigma(\omega) \label{sumrule4} . \]  The rhs has a {\it qualitatively}
different behavior in insulators and in metals. In the latter materials, in
fact, the real part of the conductivity is either finite or divergent in the
dc ($\omega \rightarrow 0$) limit, thus implying in both cases the divergence
of the integral, ergo of the ${\bf P}$ fluctuation. In insulators, instead,
the integral in the rhs of \equ{sumrule4} converges to a finite value.

Some precursor work, before SWM, attempted to relate ground-state
fluctuations to dc conductivity~\cite{Schonhammer73,Kudinov91}. It is worth
noticing that such work severely overlooks the role of BCs, while instead SWM
provide a rigorous theory in a purely transverse framework. However, SWM
neither consider different BCs, nor relate their work to the
Nozi\`eres-Pines~\cite{Nozieres58} early fluctuation-dissipation sum rule.
Here we provide a generalization of SWM to all possible BCs, \equ{sumrule3}.
Its novel outstanding message is that for any $n_\beta \neq 0$ choice the
${\bf P}$ fluctuation is finite even in metals (contrary to what stated in
Ref.~\cite{Kudinov91}).

The second cumulant moment $\langle r_\beta r_\gamma \rangle_{\rm c}$ has been
defined as a bulk property of the condensed system, which measures the quadratic
quantum fluctuations of the polarization in the many-body ground state at zero
temperature~\cite{Souza00}. It may appear therefore disturbing that its
expression, as given in \equ{sumrule3}, depends explicitly---via the $n_\beta$
coefficients---on the {\it shape} which has been chosen for taking the large-$V$
limit. In fact, this is a real physical effect and has a simple interpretation. 

The fluctuating polarization ${\bf P}$ induces a surface charge at the boundary
of the sample, which in turn generates a homogeneous depolarizing field ${\bf
E}$, which counteracts polarization: in the unperturbed ${\bf E}_0 = 0$ case
\equ{simple} reads \[ E_\beta = -4 \pi n_\beta P_\beta . \label{simple2} \] Seen
in this way, the effect obviously {\it does} depend on shape. But in condensed
matter physics one tries to steer clear from any shape issue, and therefore one
{\it interprets} \equ{simple2}, for any choice of $n_\beta$, as a choice of BCs
for performing the thermodynamic limit. Indeed, \equ{simple2} becomes the basic
one, and any reference to shape is no longer needed. The condition $\sum_\beta
{n}_\beta = 1$ is not needed either. When adopting the usual periodic BCs in all
three Cartesian coordinates, we are effectively imposing ${\bf E} = 0$, i.e.
$n_x = n_y = n_z = 0$. From what said above, and from \equ{sumrule3}, one would
obtain the same fluctuation when working in a slab geometry and addressing the
${\bf P}$ component parallel to the slab, i.e. transverse. The other extreme
case of \equ{simple2}, namely $n_x = n_y = n_z = 1$, is also well known in
condensed matter physics.  In fact, the BCs for zone-center phonon modes in
cubic binary crystals are  ${\bf E} = 0$ for transverse modes, and ${\bf E} = -
4 \pi {\bf P}$ for longitudinal ones~\cite{Huang50,Maradudin,Baroni01,note2}.

There is a complete analogy between the ground-state fluctuations of
polarization in a many-electron system at zero temperature, as discussed
here, and the equilibrium fluctuations of polarization  in a classical
dipolar system at finite temperature. In the latter case, in fact, it is well
known~\cite{Frenkel} that different BCs lead to different fluctuations but to
the {\it same} value for the static dielectric constant, provided the correct
fluctuation formula is used for each case~\cite{Neumann83}. The
shape-dependence is a combined effect of interparticle correlations and of
the long-range nature of the interactions. The analogy goes further, since
even in the quantum case the dependence on shape (or equivalently on BCs) is
a pure {\it correlation effect}, not present at the independent-electron
level (either Hartree-Fock or Kohn-Sham), where the many-body wavefunction is
a Slater determinant. In fact, the  second cumulant moment $\langle r_\beta
r_\gamma \rangle_{\rm c}$, \equ{bform}, is a function of the two-body
density: the latter, for the special case of a single-determinant
wavefunction, is an explicit function of the one-body {\it density matrix}.
As such, it can only be affected by the mean ${\bf E}$ field (i.e. zero, for
the unperturbed system), and {\it not} by its fluctuations.

This is confirmed by the present sum rule. Starting from \equ{sumrule}, we
notice that when we evaluate the rhs using the independent-electron two-body
density, we must interpret the $\alpha(\omega)$ tensor in the lhs as the
independent-electron polarizability, which  by construction neglects
self-consistency effects.  Therefore ${\bf E} = {\bf E}_0$ i.e., after
\equ{ezer}, $n_\beta = 0$.  Therefore \equ{sumrule3} reduces to
\equ{sumrule4}, which is manifestly shape-independent (or BC independent).
Incidentally, the conductivity $\sigma(\omega)$ therein must be understood as
the independent-electron conductivity. 

I now address the special form taken by \equ{sumrule2} in the longitudinal case,
where $n_x = n_y = n_z = 1$: the diagonal $zz$ component is \[ \langle z^2
\rangle_{\rm c} = - \frac{\hbar}{4 \pi^2 e^2} \frac{V}{N} \int_0^\infty \!\!\! d
\omega \; \mbox{Im }  \frac{1}{\varepsilon(\omega)} . \label{eg} \] One would
obtain the same fluctuation working in a slab geometry and addressing the
fluctuation of the ${\bf P}$ component normal to the slab. \equ{eg} applies to
correlated wavefunctions, and is invalid for independent-electron ones; it
provides a finite value both in insulators and metals. We are going to verify
the above general findings on the simplest metal of all, namely, the homogeneous
electron gas, showing that $\langle z^2 \rangle_{\rm c}$ is infinite in the
noninteracting case, and finite in the interacting one.

In order to make contact with the electron-gas literature, we need to
introduce the static structure factor defined as \[ S({\bf k}) = \frac{1}{N}
\langle \Psi | \sum_{i,j} \ei{{\bf k} \cdot ({\bf r}_i - {\bf r}_j)} | \Psi
\rangle .  \label{struct} \]  This is identically expressed in terms of the
one-- and two--body densities as: \bea S({\bf k}) & = & 1 + \frac{1}{N} \int
d {\bf r} \int d {\bf r}' \; \ei{{\bf k} \cdot ({\bf r} - {\bf r}')}
n^{(2)}({\bf r}, {\bf r}') , \nn & = & 1 + \frac{1}{N} | \tilde{n}({\bf k})
|^2  \\ \lefteqn{ + \frac{1}{N} \int \!\! d {\bf r} \!\! \int d \!\! {\bf r}'
\, \ei{{\bf k} \cdot ({\bf r} - {\bf r}')} [ \, n^{(2)}({\bf r}, {\bf r}')  -
n({\bf r}) n({\bf r'}) \, ]  .} \label{struct3} \nonumber \eea For ${\bf k} =
0$ the second term is equal to $N$, which obviously diverges in the
thermodynamic limit: such $\delta$-like singularity is neglected as usual. We
then expand in powers of ${\bf k}$ imposing centrosymmetry: therefore the
second term is quartic, and we have to second order: \bea S({\bf k}) & \simeq
& - \sum_{\alpha\beta} \frac{k_\alpha k_\beta}{2 N} \int d {\bf r} \int d
{\bf r}' \;  ({\bf r - r'})_\alpha  ({\bf r - r'})_\beta \nn & \times &[ \,
n^{(2)}({\bf r}, {\bf r}') - n({\bf r}) n({\bf r'}) \, ]  \nn & \simeq &
\sum_{\alpha\beta} \langle r_\alpha r_\beta \rangle_{\rm c} k_\alpha k_\beta
\label{struct4} , \eea Therefore for an isotropic system \[ \langle z^2
\rangle_{\rm c} = \lim_{k \rightarrow 0} \;  S(k) /k^2 . \label{limit} \] For
the noninteracting (either Hartree-Foch or Kohn-Sham) electron gas the
one-body density, and hence $S(k)$, are known exactly~\cite{Pines}: this in
fact leads to a divergent \equ{limit}. Polarization fluctuations are indeed
BC- (or shape-) independent, and diverge even in the longitudinal case, thus
confirming our general finding.

In the interacting case $S(k)$, as defined here, depends on shape via
\equ{struct4}, whereas in the existing electron-gas literature $S(k)$ is
apparently shape-independent. The reason is very simple: such literature
addresses charge fluctuations, {\it not} polarization fluctuations. It has been
stressed above that no macroscopic charge is associated to transverse
polarization fluctuations: charge fluctuations manifest themselves only within
longitudinal BCs, which are therefore {\it implicitly} assumed
by electron-gas theorists. Our longitudinal \equ{eg}, together with \equ{limit},
yields \[ S(k) \simeq - \frac{\hbar k^2}{4 \pi^2 e^2} \frac{V}{N} \int_0^\infty
\!\!\! d \omega \; \mbox{Im }  \frac{1}{\varepsilon(\omega)} , \] which indeed
is the standard fluctuation-dissipation theorem for the interacting electron
gas, known since the 1950s~\cite{Noz}. The frequency integral is finite:
replacement into \equ{limit} confirms that the longitudinal polarization
fluctuation $\langle z^2 \rangle_{\rm c}$ is finite as well.

In conclusion, I have reconciled two different forms of the
fluctuation-dissipation sum rule for quantum many-body systems: one
recent~\cite{Souza00} and one old~\cite{Noz}. The two were apparently
contradictory and apparently unrelated. Instead, I have shown that a more
general sum rule holds, yielding the previously known ones as special cases. At
the root of the generalization is a careful treatment of electron correlation in
Coulomb systems. Remarkably, the novel feature found here is a pure correlation
effect, not present at the Hartree-Fock or Kohn-Sham level.

Discussions with G. Senatore are gratefully acknowledged. Work supported by ONR
grant N00014-03-1-0570 and grant PRIN 2004 from the Italian Ministry of
University and Research.

\end{document}